# SIMULATION STUDY OF INTERNAL AND SURFACE WAVES OF VERTICALLY VIBRATED GRANULAR MATERIALS

KAI HUANG    GUOQING MIAO
PENG ZHANG    YIFEI ZHU   RONGJUE WEI

*State Key Lab of Modern Acoustics & Institute of Acoustics, Naning University, Nanjing 210093*

Molecular dynamical (MD) simulations are performed to simulate two dimensional vibrofluidized granular materials in this work. Statistics on simulation results indicate that there exist shocks propagating upward in each vibrating cycle. Under certain driving parameters surface waves similar to Faraday instability in normal fluid coexist with internal waves. Relationship between the two kinds of waves is explored. Moreover simulation results indicate that periodically structured bottom can change the dispersion relationship and amplitude of surface waves.

## 1 Introduction

Granular materials are large conglomerations of particles [1], which are very common in industry and our daily lives. Although composed of solids, they can show fluid-like behaviors as external energy input to them. Under vertical vibrations, parametric excited surface waves will appear with a dispersion relation different from that of Faraday instability in normal fluids. Other unique phenomena such as localized excitations (oscillon) [2], size and density stratification [3], cluster [4], heap formation [5] and transport [6], show the difference between this media and normal states of matter. Former studies indicate that there exist mechanical waves propagating in vibrofluidized granular materials [7]. Recently shock wave propagation inside this media is predicted by continuum equations to Navior-Stocks order and tested by molecular dynamical simulations [8] and experiments [9]. Although internal waves and surface instability are studied independently, the relationships between them are unclear until now. In this paper, we use two dimensional molecular dynamical (MD) simulations to investigate the two kinds of waves in vibrofluidized two dimensional granular materials.

Band-gap phenomena and localizations of surface waves in liquids attract more and more interest recently [10]. It has been known that periodical structure will change the dispersion relationship of parametric surface waves [11] due to Bragg resonance. We perform simulations to explore the influence of such kind of structure on surface wave of granular materials.

## 2 Method

*2.1 Model system*

A time-driven algorithm is employed to simulate $N = 2500$ hard spheres of diameter $D = 300 \mu m$ in a container of width $W = 200D$ and height $100D$. We assume that collisions are instantaneous and only binary collisions occur. At the beginning of simulation particles are uniformly distributed in the container. After every fixed time-step which is small enough to be comparable to the collision interval, we find particles that are going to collide with each other, bottom or side walls in the time step. For the particles found we update locations, translational and angular velocities according to the collision model shown below. After that, the program moves to the next time step and repeats the above process.

Normal velocities before $v_{1n}, v_{2n}$ and after $v'_{1n}, v'_{2n}$ a collision is related by

$$\begin{cases} v'_{1n} = (v_{1n} + v_{2n} + e_n \Delta v_n)/2 \\ v'_{2n} = (v_{1n} + v_{2n} - e_n \Delta v_n)/2 \end{cases} \quad (1)$$

in which the velocity related restitution coefficient is

$$e_n = \begin{cases} 1 - B \Delta v_n^\alpha & (v_n < v_0) \\ e & (v_n \geq v_0) \end{cases} \quad (2)$$

with cutoff velocity $v_0 = \sqrt{g/D}$ in which $g$ is the gravitational acceleration, and other parameters $B = 2.2$, $\alpha = 0.75$, $e = 0.75$. Surface velocities of the two particles are calculated by

$$\begin{cases} v_{1s} = w_1 r + v_{1t} \\ v_{2s} = w_2 r + v_{2t} \end{cases} \quad (3)$$

where $v_{1t}, v_{2t}$ are tangential velocities before collision, $w_1, w_2$ are angular velocities before collision, $r$ is particle radius. The surface velocities after collision are gained from

$$\begin{cases} v'_{1s} = (v_{1s} + v_{2s} + e_s \Delta v_s)/2 \\ v'_{2s} = (v_{1s} + v_{2s} - e_s \Delta v_s)/2 \end{cases} \quad (4)$$

in which the restitution coefficient $e_s$ obey the same rule as $e_n$. Tangential and angular velocities $v'_{1t}, v'_{2t}, w'_{1t}, w'_{2t}$ after collision are,

$$\begin{cases} v'_{1t} = v_{1t} + \delta(v'_{1s} - v_{1s}) \\ v'_{2t} = v_{2t} + \delta(v'_{2s} - v_{2s}) \end{cases} \quad (5)$$

$$\begin{cases} w'_{1t} = w_1 + (1-\delta)(v'_{1s} - v_{1s})/r \\ w'_{2t} = w_2 + (1-\delta)(v'_{2s} - v_{2s})/r \end{cases} \quad (6)$$

in which $\delta = 0.28$. The plate undergoes sinusoidal vibration with two control parameters, frequency $f$ and normalized vibration acceleration $\Gamma = 4\pi^2 f^2 A/g$ where $A$ is the vibration amplitude. Restitution coefficient of collisions of spheres with container bottom and side walls in the normal direction obeys the same rule as particle-particle collisions and in the colliding surface the restitution coefficient is $1$.

*2.2 Statistics of simulation results*

The simulation results are recorded with a sample rate $N_p f_s$ where $N_p = 50$. The space of every frame is divided into $2d \times d$ cells with left boundary and plate starting points. We choose $f_s = f/2$ because the parametric excitation waves are subharmonic. Assembly average is performed over $100$ cycles for granular properties at $N_p$ phase points in each acquisition cycle. Phase time $0$ corresponds to the time when the bottom is moving upward with maximum velocity. Granular density $D(i,j,k)$ is the number of particles in the cell located in the $ith$ column and $jth$ layer at phase time $k$. Granular temperature is defined by

$$T(i,j,k) = \sum_{n=1}^{N_c} |\vec{v}_n - \vec{v}_b(i,j,k)|^2 / 2N_c, \quad (7)$$

in which $N_c$ is the total number of particles in cell $(i,j)$ at time $k$, $\vec{v}_n$ is the velocity of the $nth$ particle and the background velocity is

$$\vec{v}_b(i,j,k) = \sum_{n=1}^{N_c} \vec{v}_n / N_c \quad (8)$$

Time-space dependent Mach number is calculated by:

$$Ma = \left| \frac{v_{bn} - v_p}{c} \right|, \quad (9)$$

in which $v_{bn}$ is the vertical component of $\vec{v}_b$, $v_p$ is the velocity of the plate and $c$ is

the speed of sound in the media calculated with [8,13]

$$c = \sqrt{T\chi(1+\chi+\frac{v}{\chi}\frac{\partial \chi}{\partial v})},\qquad(10)$$

where $\chi = 1 + 2(1+e)v[1-(v/v_m)^{4v_m/3}]^{-1}$. Here $e$ is the restitution coefficient, $v$ is the volume fraction and $v_m$ is the close packed volume fraction.

**3 Results and discussion**

*3.1 Shock propagation in granular layer with surface instability*

In each vibration cycle, granular layer will leave the plate as nondimensional acceleration of the vibrating plate less than $-1$. Later in that cycle, when the plate moves upward and collides with the bottom of the granular layer, a high density as well as a shock forms. The shock is identified in the region where the Mach number increases from near zero to a value greater than unity. From two dimensional granular properties at different times in one sampling period, we can see the formation of surface instability together with the propagation of shocks in the media. After about 20 cycles of vibration, surface instability with wavelength $\lambda = 46d$ appears. Here we show results with one wavelength from $L = 30$ to 55. At $time = 1.36\pi$, the distance between granular layer and the plate reaches a maximum. The layer expands as moving downward freely, which leads to the increase of the amplitude of the surface wave. A shock is found with the Mach number increases from nearly zero at $z = 10$ to $Ma = 3$ at $z = 17$. In the following free flight time, the layer continuously expands the shock propagates through granular layer and disappears. After the height of the instability reaches its maximum, granular layer compresses due to collision with the plate at $time = 1.84\pi$. Velocity field at this time shows that particles at $z = 1$ collide with the plate and move upward with the plate while particles at height $z \geq 3$ still fly downward. In the region $1 \leq z \leq 3$ the Mach number increases to be greater than unity indicating that a new shock region appears near the plate. At $time = 2.00\pi$ the shock propagates upward to $z = 6$ and stops at that height. Collisions between downward flowing particles at $z > 6$ and upward flowing particles at $z < 6$ lead to horizontal movements of these particles. With the collisions vertical kinetic energy is transferred into that in horizontal direction. As time elapses, the amplitude of surface instability decreases until disappears at $time = 2.40\pi$. Horizontal movements of particles lead to the appearance of an area where horizontal collisions occur frequently at $L = 53$ which is $\lambda/2$ away from the disappeared instability and the center of the following peak. At $time = 2.40\pi$, granular layer leaves the plate and flies freely again. New surface wave appears with a shift of peak locations of half a wavelength. With the new surface wave, a new shock forms and propagates upward with the maximum Mach number increases. At $time = 3.36\pi$, granular layer reaches its maximum height, shock propagates to $z = 17$ and a new vibration cycle begins. The above results agree with our former experiments that there exist shock propagation upward in one vibration cycle [9]. However, simulation results here indicate that there exist two shocks in one vibration cycle. One forms with the collision between granular layer and the plate and stops propagating at some height where vertical movements transfer to horizontal movements. This results in the formation of surface instability. The other forms with the formation of surface instability and propagates upward until the top of granular layer.

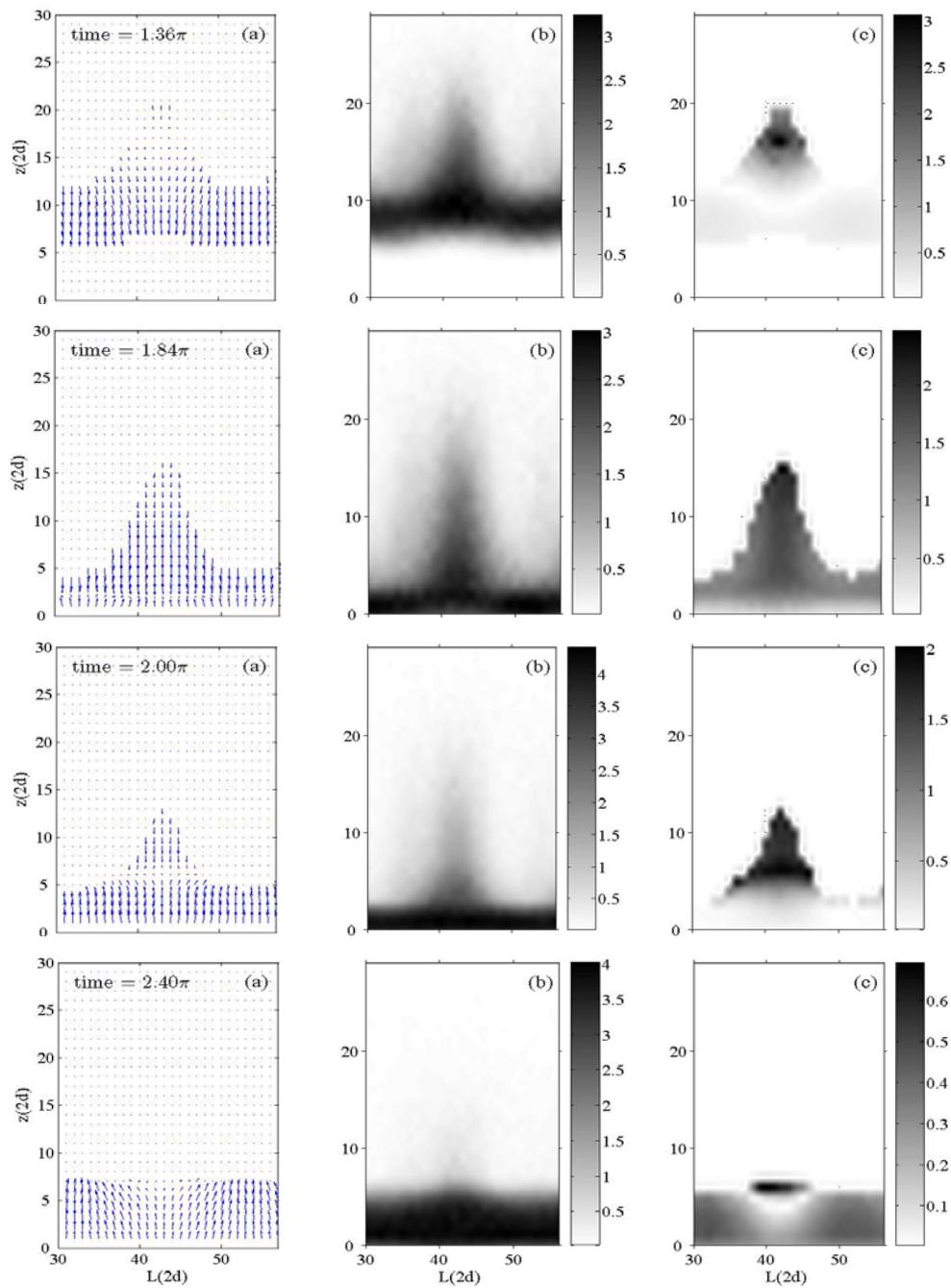

Fig.1 Space distributions of velocity vector (a), density (b) and Mach number (c) at four times, in which (b) and (c) are shaded according to values at corresponding cells. Density means the average particle number in each cell. Control parameters are $\Gamma = 4$ and $f = 20$.

*3.2 Influence of periodical bottom structure on surface waves*

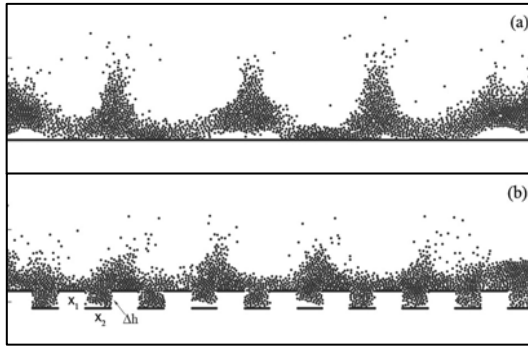 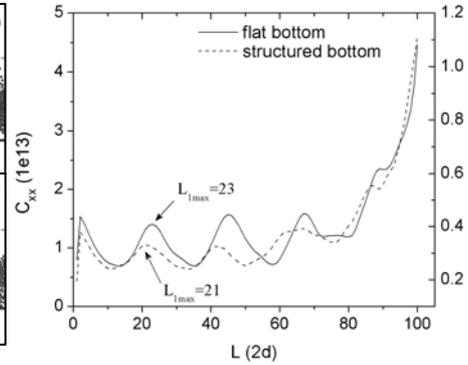

Fig.2 Snapshots of simulation results without (a) and with (b) periodical bottom structures at phase time $1.84\pi$ when the flying period ends and surface instability reaches its maximum height. Driving parameters are $f = 20Hz, \Gamma = 3$.

Fig.3 Results of particle-particle correlation results as a function of $L$. Result of flat bottom is shown in the left-bottom axes and that of structured bottom is shown in the right-bottom axes.

In former simulation and experimental results of surface instability in vibrofluidized granular layer an empirical dispersion relationship $\lambda = \sqrt{H}(\lambda_0 + g_{eff}/f^2)$ is proposed [14], in which $H$ is the layer depth in the unit of particle diameter, $\lambda_0$ and $g_{eff}$ are two control parameters. The dispersion relationship indicates that periodical surface instability will be influenced by a bottom with a periodical structure. Therefore we perform simulations with periodical bottom topography. The period of the bottom structure is $20d$ with $x_1 = x_2 = 10d$ and $\Delta h = 8d$ (as shown in Fig.2(b)). The wavelength of surface instability is measured by a particle-particle correlation function,

$$C_{xx}(L) = (1/(W-L)N^2)\sum_{i=1}^{N}\sum_{j=1}^{N}\delta[L-|L_i-L_j|] \tag{11}$$

in which

$$\delta[L] = \begin{cases} 1 & L=0 \\ 0 & L \neq 0 \end{cases}. \tag{12}$$

The location of the $nth$ peak value corresponds to the $n\lambda$ of the surface wave. Comparison of simulation results (shown in Fig.2 (a) and (b)) indicates that the amplitude of surface wave decreases as the structured bottom is added. Under the same environmental parameters, the wavelength of surface instability with flat bottom ($46d$) is greater than that of surface instability with structured bottom ($42d$). Results of velocity vector shown in *section 3.1* indicate that horizontal movements of particles at the free flight period of granular layer are critical for the formation of surface instability. Thus the reduction of the amplitude of surface instability in Fig.2 (b) is due to that horizontal movements of particles are inhibited by the periodical structure. The decrease of wavelength indicates that the dispersion relationship of vibrofluidized granular materials changes due to the bottom structure. Compare this result with that in normal fluids, we may conclude that Bragg resonance caused by the periodical structure is responsible for the change of dispersion relationship.

**4 Conclusions**

In conclusion, the internal shock wave as well as the surface instability is explored in this paper. When granular layer collides with the plate, a shock region which is defined as the increase region of Mach number from lower to higher than unity is found to form and propagate upward as the plate moves upward and granular layer compresses. The region stops propagating when it reaches a critical height. At the same time, collisions in this region lead to horizontal movements of particles. Horizontal collisions at the free flight period result in the formation of new surface wave with its peaks half a wavelength away from the former surface instability and the other shock which propagates upward to the top of granular layer.

Moreover, we perform simulations with structured bottom and compare the results with those of flat bottom. The height $\Delta h$ of the bottom structure is found to take responsibility for the inhibition of the amplitude of surface instability and the period of the bottom may be the reason for the change of dispersion relationship.

## 5 Acknowledgments

This work was supported by the Special Funds for Major State Basic Research Projects, National Natural Science Foundation of China through Grant No. 10474045 and No. 10074032, and by the Research Fund for the Doctoral Program of Higher Education of China under Grant No. 20040284034.

The numerical simulation in this paper is accomplished on high performance computer, sgi origin 3800, of Nanjing University.